\documentclass[conference]{IEEEtran}
\IEEEoverridecommandlockouts 

\usepackage{citesort,amsbsy,amsthm}
\ifCLASSINFOpdf
  \usepackage[pdftex]{graphicx}
  \graphicspath{{../pdf/}{../jpeg/}}
  \DeclareGraphicsExtensions{.pdf,.jpeg,.png}
\else
  \usepackage[dvipdfmx]{graphicx}
  \graphicspath{{../eps/}}
  \DeclareGraphicsExtensions{.eps}
\fi
\usepackage{amssymb,array,mdwmath,mdwtab}
\usepackage{enumerate,subfig}
\usepackage{multicol,multirow,paralist}
\usepackage{color}
\usepackage{algorithm}
\usepackage{algorithmic}

\usepackage[utf8]{inputenc} 
\usepackage[T1]{fontenc}
\usepackage{url}
\usepackage{ifthen}
\usepackage{cite}
\usepackage[cmex10]{amsmath} 

\begin{document}

\title{Channel Decoding with Quantum Approximate Optimization Algorithm}

\author{
  \IEEEauthorblockN{
    Toshiki Matsumine\IEEEauthorrefmark{3}\IEEEauthorrefmark{2}\thanks{T. Matsumine conducted this research when he was an intern at MERL.},
    Toshiaki Koike-Akino\IEEEauthorrefmark{3},
    and Ye Wang\IEEEauthorrefmark{3}
  }
  \IEEEauthorblockA{
    \IEEEauthorrefmark{3}
    Mitsubishi Electric Research Laboratories (MERL), 
    201 Broadway, Cambridge, MA 02139, USA. \\
    \IEEEauthorrefmark{2}
    Department of Electrical and Computer Engineering,
    Yokohama National University, \\
    79-5 Tokiwadai, Hodogaya, Yokohama, Kanagawa, Japan\\
    Email: matsumine-toshiki-tk@ynu.jp,
    \{koike, yewang\}@merl.com
  }
}

\maketitle

\begin{abstract}
Motivated by the recent advancement of quantum processors, we investigate quantum approximate optimization algorithm (QAOA) to employ quasi-maximum-likelihood (ML) decoding of classical channel codes. 
QAOA is a hybrid quantum-classical variational algorithm, which is advantageous for the near-term noisy intermediate-scale quantum (NISQ) devices, where the fidelity of quantum gates is limited by noise and decoherence.
We first describe how to construct Ising Hamiltonian model to realize quasi-ML decoding with QAOA.
For level-1 QAOA, we derive the systematic way to generate theoretical expressions of cost expectation for arbitrary binary linear codes.
Focusing on (7, 4) Hamming code as an example, we analyze the impact of the degree distribution in associated generator matrix on the quantum decoding performance.
The excellent performance of higher-level QAOA decoding is verified when Pauli rotation angles are optimized through meta-heuristic variational quantum eigensolver (VQE).
Furthermore, we demonstrate the QAOA decoding performance in a real quantum device.
\end{abstract}


%
\IEEEpeerreviewmaketitle

\section{Introduction}

Quantum computers can offer a significant potential to accomplish more efficient computations compared to traditional digital computers for various problems by exploiting quantum-mechanism, e.g., superposition, entanglement, and quantum tunneling, in terms of not only execution time but also energy consumption. It is highly expected that quantum computers could provide breakthroughs in wide range of research domain, such as chemical engineering, complex system optimizations, and artificial intelligence. Not only theoretical concept, several commercial quantum computers have been already built by several industries including IBM, google, honeywell, and so on. For instance, IBM has made $5$ and $16$ qubits quantum computers available to the public via cloud service.

Quantum error correction for protecting quantum states from undesired noise and decoherence has been investigated in literature, e.g., \cite{calderbank1996good,calderbank1998quantum,poulin2009quantum,mackay2004sparse,wilde2013polar}. Against these previous works, in this paper, we consider decoding {\em {classical}} binary linear codes assisted by quantum computing.
Since the maximum-likelihood (ML) decoding of channel codes, which has non-deterministic polynomial-time hardness \cite{berlekamp1978inherent}, is feasible only when the block length is very short, the most popular approach in the communication standards is based on sub-optimal belief-propagation (BP) algorithm combined with probabilistic codes, such as turbo codes and low-density parity-check (LDPC) codes. Although these approaches already give the practical solutions, the decoding complexity may be further increasing when we consider improving the performance of channel codes with finite block lengths and finite iterations. For example, nonbinary codes has been shown to achieve significant performance gain over binary counterparts in the short-to-medium block length regime at the cost of increasing decoding complexity. Channel decoding with hybrid quantum-classical algorithm can be a potential new framework to significantly reduce the decoding complexity for those scenarios.

Applications of quantum algorithm to communication systems have been investigated, e.g., in \cite{babar2015exit,botsinis2015iterative,botsinis2018quantum}. In \cite{botsinis2015iterative}, the quantum-assisted iterative detection for multi-user systems was proposed. Joint channel estimation and data detection was studied in the uplink of multi-input multi-output (MIMO) systems \cite{botsinis2016joint}. Optimization of vector perturbation precoding for the multi-user transmission was proposed in \cite{botsinis2016quantum}. However, they typically assume that long qubits (e.g., more than 1000) are available and that quantum-circuit gates have no errors, which may be beyond the capabilities of near-term noisy intermediate-scale quantum (NISQ) computers.

A hybrid quantum-classical algorithm called quantum approximate optimization algorithm (QAOA) was recently proposed to solve various NP-hard problems \cite{edward2014qaoa}.
Because of high robustness against quantum errors, QAOA is expected to be a suitable candidate for NISQ devices \cite{patrick2018quantum} and a breakthrough driver towards \textit{quantum supremacy} \cite{edward2016quantum}. 
For level-$p$ QAOA, classical discrete optimization can be probabilistically solved by mapping to an Ising Hamiltonian with $2p$ annealing parameters.
The approximation quality of QAOA improves towards the ground state when increasing $p$. Even for $p=1$, QAOA has guaranteed better probable performance than classical algorithms for certain problems such as MaxCut.
In order to optimize QAOA ansatz, we typically use variational quantum eigensolver (VQE) where a quantum processor performs an expectation calculation of the cost function and the $2p$ parameters are optimized by a classical computer in a closed loop. 

The main contributions of this paper are summarized below:
\begin{itemize}
 \item We develop a new framework with QAOA quantum processor to decode classical channel codes;
 \item We introduce an Ising Hamiltonian for QAOA to realize quasi-ML decoding of arbitrary linear binary codes;
 \item We propose a theoretical method to derive quantum eigenvalue expressions for level-1 QAOA decoding;
 \item We analyze the impact of the degree distribution of (7, 4) Hamming codes on quantum decoding performance;
 \item We implement QAOA decoder with both quantum simulators and real quantum computers.
\end{itemize}


\section{Preliminaries}

\subsection{Quantum Bit (Qubit)}

\label{sec:pre}
In quantum systems, a {\em{qubit}} is expressed as the following state superposing bases of $|0\rangle$ and $|1\rangle$:
$| \phi \rangle = \alpha_0 |0\rangle + \alpha_1 |1\rangle$,
where $\alpha_1$ and $\alpha_2$ are complex numbers subject to $|\alpha_0|^2+|\alpha_1|^2=1$.
When qubits are measured, the classical bit $0$ or $1$ is observed with a probability of $|\alpha_0|^2$ or $|\alpha_1|^2$, respectively.
The above \textit{ket-notation} corresponds to column-vector operations of the two basis states $|0\rangle=[1,0]^\mathrm{T}$ and $|1\rangle=[0,1]^\mathrm{T}$, whereas
the \textit{bra-notation} is used for row-vector operations corresponds to its Hermitian transpose; i.e., 
$\langle \phi| = |\phi \rangle^\dag = [\alpha^*_0, \alpha_1^*]$. Here, $[\cdot]^\dag$, $[\cdot]^*$ and $[\cdot]^\mathrm{T}$ denote Hermitian transpose, 
complex conjugate and transpose, respectively.
Note that multi-qubit state is represented by sum of Kronecker products of basis vectors such as $|000\rangle = |0\rangle^{\otimes 3}$.

\subsection{Quantum Gates}
The basic operations on a qubit is defined as a unitary matrix, which is called {\em{gate}}. Some of the most common gates are associated with Pauli matrices:
$\mathbf{I} = \left[\begin{smallmatrix}
                     1 & 0\\
                     0 & 1
                    \end{smallmatrix}\right]$,
$\mathbf{X} = \left[\begin{smallmatrix}
                     0 & 1\\
                     1 & 0
                    \end{smallmatrix}\right]$,
$\mathbf{Y} = \left[\begin{smallmatrix}
                     0 & -\jmath\\
                     \jmath & 0
                    \end{smallmatrix}\right]$, and
$\mathbf{Z} = \left[\begin{smallmatrix}
                     1 & 0\\
                     0 & -1
                    \end{smallmatrix}\right]$,
where $\jmath=\sqrt{-1}$ is the imaginary unit.
The X gate is bit-flip (i.e., NOT operation), Z gate is phase-flip, and Y gate flips both bit and phase.
%
The Hadamard (H) gate is used to generate a superposition state $|+\rangle = \tfrac{1}{\sqrt{2}} |0\rangle + \tfrac{1}{\sqrt{2}} |1\rangle$:
$\mathbf{H} = \tfrac{1}{\sqrt{2}}
\left[
\begin{smallmatrix}
1 & 1 \\
1 & -1
\end{smallmatrix}
       \right]$.
A controlled-NOT (CNOT or CX) gate is a multi-qubit gate that flips the target qubit if and only if the control qubit is $| 1 \rangle$:
\begin{align}
\mathsf{CNOT} =
 \left[
\begin{smallmatrix}
1 & 0 & 0 & 0\\
0 & 1 & 0 & 0\\
0 & 0 & 0 & 1\\
0 & 0 & 1 & 0
\end{smallmatrix}
 \right]
.
\end{align}

\subsection{QAOA Algorithm}
\label{sec:qaoa}
QAOA algorithm\cite{edward2014qaoa} was proposed for discrete optimization problems, such as the MaxSat, MaxCut, and MaxClique, which are expressed as an unconstrained discrete optimization:
\begin{align}
C(\mathbf{z}) = \sum_{\nu=1}^{n} C_\nu(\mathbf{z}),
\end{align}
where $\mathbf{z} = [z_1, z_2, \ldots, z_k] \in \mathbb{F}^k_2$ denotes binary label, and $C_\nu(\mathbf{z})$ is the $\nu$th binary function to satisfy, which are called {\em {clause}}. In this problem formulation, we try to find a binary vector $\mathbf{z}$ that maximizes the number of satisfied clause $C_\nu(\mathbf{z})$.

For typical QAOA, quantum state is initialized to admixing superposition state $|+\rangle^{\oplus k}$, which produces equi-probable random bits $\mathbf{z}$ once measured when no other operations. Letting $U(C, \gamma)$ denote a unitary operator for the cost Hamiltonian $C$ with an angle $0 \leq \gamma \leq 2\pi$, which is defined as 
\begin{align}
U(C, \gamma) = \exp(-\jmath \gamma C) = \prod^{n}_{\nu=1} \mathrm{e}^{-\jmath \gamma C_\nu}.
\end{align}
Consider a driver Hamiltonian defined by $B=\sum^k_{\kappa=1} \mathbf{X}_\kappa$, which flips $k$-qubits independently. We use the following unitary operator with an angle $0 \leq \beta \leq \pi$:
\begin{align}
U(B, \beta) = \exp(-\jmath \beta B) = \prod^{k}_{\kappa=1} \mathrm{e}^{-\jmath \beta \mathbf{X}_\kappa}.
\end{align}
Note that this driver Hamiltonian has the ground-state eigenvector of $|\phi\rangle = |+\rangle^{\otimes k}$.

QAOA uses alternating quantum operator ansatz circuits of depth $p$ based on Hamiltonians $B$ and $C$ to maximize the expected cost function, with $2p$ angle parameters $\boldsymbol{\gamma}$ and $\boldsymbol{\beta}$:
\begin{align}
|{\boldsymbol{\gamma}, \boldsymbol{\beta}}\rangle =
U(B, \beta_p)U(C, \gamma_p) \cdots U(B, \beta_1)U(C, \gamma_1) |\phi\rangle.
\end{align}
Let $F_p$ denote the expectation of the cost function $C$ as
\begin{align}
F_p({\boldsymbol{\gamma}, \boldsymbol{\beta}}) =
 \langle C \rangle(\boldsymbol{\gamma}, \boldsymbol{\beta})
 =\langle {\boldsymbol{\gamma}, \boldsymbol{\beta}} | C | {\boldsymbol{\gamma}, \boldsymbol{\beta}} \rangle,
\end{align}
and let $F^\star_p$ be the maximum of $F_p({\boldsymbol{\gamma}, \boldsymbol{\beta}})$ over the angles:
$F^\star_p = \max_{{\boldsymbol{\gamma}, \boldsymbol{\beta}}} F_p({\boldsymbol{\gamma}, \boldsymbol{\beta}})$.
The objective of QAOA algorithm is to maximize $F^\star_p$ by properly choosing parameters $\boldsymbol{\gamma}, \boldsymbol{\beta}$.
The quality of the approximation improves as $p$ increases and the global optimum maximizing cost function $C(\mathbf{z})$ can be asymptotically achieved when infinite depth $p$, i.e.,  $\lim_{p \rightarrow \infty} F^\star_p = \max_\mathbf{z} C(\mathbf{z})$.

In QAOA, the calculation of the expectation $F_p({\boldsymbol{\gamma}, \boldsymbol{\beta}})$ is performed by repeated measurements with quantum computers based on variational principle in the computational basis, which is infeasible for classical computers as $p$ increases. The search for optimal variational parameters ${\boldsymbol{\gamma}, \boldsymbol{\beta}}$ that maximize $F_p({\boldsymbol{\gamma}, \boldsymbol{\beta}})$ are efficiently performed by classical computers, e.g., employing Nelder--Mead (NM) method\cite{edward2016quantum}.

\section{Channel Decoding with QAOA Algorithm}
\label{sec:dec}

Here, we propose to use QAOA for performing quasi-ML decoding of linear
error-correcting codes for digital communications over noisy classical
channels.

\subsection{Digital Communications Model}
We consider an ($n$, $k$) binary linear code, specified by a generator matrix of $\mathbf{G} \in \mathbb{F}_2^{k\times n}$, where $n$ and $k$ are the codeword bit length and information bit length, respectively.
The codeword $\mathbf{x} \in \mathbb{F}^n_2$ is generated as
 $\mathbf{x} =  \mathbf{u}\mathbf{G}$,
where arithmetic operation based on modulo-2 is taken place and $\mathbf{u} \in \mathbb{F}^k_2$ is the binary information vector.
Over digital transmission channels, the received signal is modeled as
$\mathbf{y} = \mathbf{x} + \mathbf{w}$,
where $\mathbf{y}\in \mathbb{F}^n_2$ and $\mathbf{w} \in \mathbb{R}^n$ are the received and noise vectors, respectively.
In this paper, we focus on binary-symmetric channel (BSC) for simplicity since extension to other channels such as additive white Gaussian noise channel is straightforward.

For BSC channels, the problem of channel decoding is to find the codeword such that the Hamming distance from the received signal is minimized, i.e.,
\begin{align}
  \arg\min_{\mathbf{u}} d_\mathrm{H}(\mathbf{y}|\mathbf{x}) = 
 \arg\max_{\mathbf{u}} \sum^{n}_{\nu=1} (1-2y_\nu) (1-2x_\nu),
  \label{eq:cost}
\end{align}
where $d_\mathrm{H}(\mathbf{y}|\mathbf{x})$ is the number of elements of two vectors $\mathbf{x}$ and $\mathbf{y}$ which differ. This is equivalent to maximizing the correlation between the transmitted codeword and the received vector.

\subsection{Construction of Cost Hamiltonian}
The proposed quantum decoder operates on $k$-qubit space corresponding to information bits $\mathbf{u}$. The objective of quantum decoding is to find most-likely $k$-qubit states that maximize \eqref{eq:cost}. To do so, we consider the following cost Hamiltonian:
\begin{align}
  C = \sum^n_{\nu=1} C_\nu = \sum^n_{\nu=1} (1-2y_\nu) \prod_{\kappa \in \mathcal{I}^\mathrm{c}_\nu} \mathbf{Z}_\kappa,
  \label{eq:ham}
\end{align}
where $\mathcal{I}^\mathrm{c}_\nu$ is a set of nonzero-element indices in the $\nu$th column of a generator matrix $\mathbf{G}$, i.e., $\mathcal{I}^\mathrm{c}_\nu=\{\kappa: [\mathbf{G}]_{\kappa,\nu}=1 \}$ where $[\cdot]_{i,j}$ denotes the element of an argument matrix at the $i$th row and $j$th column. Since the Z-gate performs as $+|\phi\rangle$ or $-|\phi\rangle$ for the qubit state of $|\phi\rangle = |0\rangle$ or $|1\rangle$, respectively, maximizing the cost Hamiltonian \eqref{eq:ham} is equivalent to ML decoding \eqref{eq:cost}.

\subsection{Degree Optimization of Generator Matrix}
The proposed cost Hamiltonian in \eqref{eq:ham} is a generalized version used for MaxCut problem\cite{edward2014qaoa}, in which case the column degree is identical to be two, i.e., $d^\mathrm{c}_\nu = |\mathcal{I}^\mathrm{c}_\nu| = 2$, regardless of column $\nu$.
Even for such regular degree-2 cost Hamiltonian, it was shown in \cite{hadfield2018qaoa} that the quality of QAOA approximation is highly dependent on the graph connectivity, specifically, the number of girth-6 (i.e., triangles in graph) and row degrees $d^\mathrm{r}_\kappa = |\mathcal{I}^\mathrm{r}_\kappa|$ for a row-wise nonzero-entry index set of $\mathcal{I}^\mathrm{r}_\kappa = \{\nu: [\mathbf{G}]_{\kappa, \nu} = 1\}$.

In order to obtain an insight to optimize generator matrix $\mathbf{G}$ suited for our QAOA decoding, we consider to create different degree distributions by applying the following basis transform to the original generator matrix $\mathbf{G}$:
$ \mathbf{G}' = \mathbf{P} \mathbf{G}$,
where $\mathbf{P} \in \mathbb{F}_2^{k\times k}$ is a full-rank matrix that performs basic row operations.
It should be noticed that the performance of linear block codes over symmetric channels is invariant with respect to such a basis transform for the classic ML decoder because the Hamming weight spectrum remains unchanged. In the case of QAOA decoder, however, the decoder performance depends on the specific structure of generator matrices.

For example, the generator matrix of (7, 4) systematic Hamming codes is given by
\begin{align}
\mathbf{G} =
\left[
\begin{smallmatrix}
1 & 0 & 0 & 0 & 1 & 1 & 0\\
0 & 1 & 0 & 0 & 1 & 0 & 1\\
0 & 0 & 1 & 0 & 0 & 1 & 1\\
0 & 0 & 0 & 1 & 1 & 1 & 1
\end{smallmatrix}
 \right]
.
 \label{eq:G}
\end{align}
The column-degree distribution of this matrix is $[1, 1, 1, 1, 3, 3, 3]$, whose average degree is $\bar{d}^\mathrm{c} = \mathbb{E}[ d^\mathrm{c}_\nu] = 1.86$ with $\mathbb{E}[\cdot]$ denoting an expectation.
Suppose the following transform matrix for instance:
\begin{align}
\mathbf{P} =
 \left[
\begin{smallmatrix}
1 & 1 & 1 & 1\\
0 & 1 & 1 & 1\\
1 & 0 & 1 & 0\\
0 & 0 & 0 & 1
\end{smallmatrix}
 \right]
,
\end{align}
then the new generator matrix will be written as
\begin{align}
\mathbf{G}' =
 \left[
\begin{smallmatrix}
1 & 1 & 1 & 1 & 1 & 1 & 1\\
0 & 1 & 1 & 1 & 0 & 0 & 1\\
1 & 0 & 1 & 0 & 1 & 0 & 1\\
0 & 0 & 0 & 1 & 1 & 1 & 1
\end{smallmatrix}
 \right]
,
\end{align}
whose average degree increases to $\bar{d}^\mathrm{c} = 2.71$ without changing Hamming weight spectrum of the linear codes.
In this way, we can evaluate the performance of QAOA decoder with various generator matrices to optimize degree distributions.

%
%

\section{Performance Analysis of QAOA Decoding}

We describe how to systematically analyze the performance of QAOA channel decoding. We generalize the analytical method\cite{hadfield2018qaoa} investigated for MaxCut problem, by considering irregular degree distributions. We focus on level-1 QAOA having single driver $B$ and cost $C$ Hamiltonians. 

\subsection{Theoretical Analysis Method for Level-1 QAOA}

For simplicity of analysis in idealistic QAOA behavior, we here assume
zero-word transmission over error-free channels
$\mathbf{y}=\mathbf{x}=\mathbf{0}$ as generalization is
straightforward. Since the cost expectation can
be decomposed as $F_p(\boldsymbol{\gamma}, \boldsymbol{\beta}) =
\sum_\nu \langle C_\nu \rangle$, we focus the $\nu$th cost Hamiltonian
whose degree is $d_\nu^\mathrm{c} = |\mathcal{I}_\nu^\mathrm{c}|$. For
example in Hamming code generator \eqref{eq:G}, the 5th column
$[\mathbf{G}]_{:,5} = [1,1,0,1]^\mathrm{T}$ corresponds to the
Hamiltonian $C_5 = (1-2y_5) \mathbf{Z}_1 \mathbf{Z}_2 \mathbf{Z}_4$
whose degree is $d^\mathrm{c}_5=3$.

In $\langle C_\nu \rangle = \langle \phi| U^\dag(C, \gamma_1) U^\dag(B,
\beta_1) C_\nu U(B, \beta_1) U(C, \gamma_1) |\phi \rangle$, observe that most terms in the operator $U(B, \beta_1) = \prod \exp(-\jmath \beta_1 \mathbf{X}_i)$ will commute and result in
\begin{align}
 &U(B, \beta_1)^\dag \bigl( \prod \mathbf{Z}_\kappa \bigr) U(B, \beta_1)
 =
 \prod
 (c' \mathbf{Z}_\kappa + s' \mathbf{Y}_\kappa),
 \label{eq:BC}
\end{align}
where $c'=\cos(2\beta_1)$ and $s'=\sin(2\beta_1)$, by recalling that
Pauli matrix $\mathbf{\Sigma} \in \{\mathbf{X}, \mathbf{Y},
\mathbf{Z}\}$ satisfies $\exp(\jmath \beta \mathbf{\Sigma}) =
\cos(\beta) \mathbf{I} + \jmath \sin(\beta) \mathbf{\Sigma}$ and
circulation rule such as $\mathbf{X} \mathbf{Z} = -\jmath \mathbf{Y}$.
The $d^\mathrm{c}_\nu$-ary product of binary additions in \eqref{eq:BC}
can be expanded in the $2^{d_\nu^\mathrm{c}}$-ary sum of
$d_\nu^\mathrm{c}$-ary products.  Let $\mathbf{b} \in \mathbb{F}_2^k$
represent such terms to indicate either $c'\mathbf{Z}_\kappa$ or
$s'\mathbf{Y}_\kappa$ is used at non-zero locations of
$[\mathbf{G}]_{:,\nu}$. For example, the Hamiltonian $C_5$ in
\eqref{eq:G} needs to account for $2^3$-terms of $c'^3 \mathbf{Z}_1
\mathbf{Z}_2 \mathbf{Z}_4$, $c'^2 s' \mathbf{Z}_1 \mathbf{Z}_2
\mathbf{Y}_4$, $\ldots$, $s'^3 \mathbf{Y}_1 \mathbf{Y}_2 \mathbf{Y}_4$
by associative binary vector of $\mathbf{b} = [0,0,0,0]$, $[0,0,0,1]$,
$\ldots$, $[1,1,0,1]$, respectively. Letting $\varpi$ be the weight of
binary vector $\mathbf{b}$, the cost expectation will be proportional to
$(-\jmath s')^{\varpi} c'^{(d^\mathrm{c}_\nu - \varpi)}$ due to
$\mathbf{Z} \mathbf{Y} = -\jmath \mathbf{X}$ and $\langle + | \mathbf{X}
| + \rangle = 1$.

Next, we consider cost operator $U(C, \gamma_1)$ on each
decomposed Pauli terms $\mathbf{W}^\mathbf{b}$. Selecting only non-commutable cost Hamiltonians $C^{\mathbf{b}}$, we can write
\begin{align}
 U(C, \gamma_1)^\dag \mathbf{W}^\mathbf{b}  U(C, \gamma_1)
 &=
 U(C^\mathbf{b}, 2\gamma_1)^\dag \mathbf{W}^\mathbf{b}.
\end{align}
We refer to the number of such non-commutable Hamiltonians as the rank
$\rho$. It can be obtained by selecting columns of $\mathbf{G}$ having
odd weight after Hadamard product of $\mathbf{b}$. For example, $\mathbf{b} = [1,1,0,1]$ corresponding to
$\mathbf{W}^\mathbf{b} = s'^3 \mathbf{Y}_1\mathbf{Y}_2\mathbf{Y}_4$ yields
$\mathbf{1}^\mathrm{T} (\mathbf{G} \odot \mathbf{b}) = [1, 1, 0, 1, 3,
2, 2]$, and thus non-commutable sub-matrix is
$\mathbf{G}^\mathbf{b} = [\mathbf{G}]_{:,\{1,2,4,5\}}$ whose rank is
$\rho=4$.

We then consider binary representation of combinatorials in
\begin{align}
 &
 U(C^\mathbf{b}, 2\gamma_1)^\dag = \prod^\rho_\nu \mathrm{e}^{2\jmath \gamma_1 C_\nu} 
 =
 \prod^\rho_\nu \bigl(c \mathbf{I} + \jmath s \prod_\kappa \mathbf{Z}_\kappa \bigr),
\end{align}
where $c=\cos(2(-1)^y\gamma_1)$ and
$s=\sin(2(-1)^y\gamma_1)$. Specifically, letting $\mathbf{a} \in
\mathbb{F}_2^\rho$ indicate binary choice of either $c\mathbf{I}$ or
$\jmath s\prod \mathbf{Z}_\kappa$, the above $\rho$-ary products of binary
additions can be expressed by $2^\rho$-ary sums of $\rho$-ary
multiplications. For example, $(c \mathbf{I})(c \mathbf{I})(\jmath s
\mathbf{Z}_1 \mathbf{Z}_2 \mathbf{Z}_4)(\jmath s \mathbf{Z}_1
\mathbf{Z}_3 \mathbf{Z}_4) = - s^2 c^2 \mathbf{Z}_2 \mathbf{Z}_3$
corresponds to $\mathbf{a}=[0,0,1,1]$ for sub-generator
$[\mathbf{G}]_{:,\{1,2,4,5\}}$. Letting a weight of binary vector
$\mathbf{a}$, the cost function will be proportional to $(\jmath
s)^\omega c^{(\rho-\omega)}$ if it is non-commutable to cost Hamiltonian
associated with $\mathbf{b}$. Specifically, such $\mathbf{a}$ must hold
$\mathbf{b} = \mathbf{G}^\mathbf{b} \mathbf{a}$. There may exist plural
of such binary vector pairs $\mathbf{a}$ and $\mathbf{b}$. Define
$A_\nu^{\mathbf{a}, \mathbf{b}}$ as the number of such pairs.

In consequence, the cost expectation can be obtained by counting the number of
combinatorials subject to $\mathbf{b} = \mathbf{G}^\mathbf{b} \mathbf{a}$
for each column $\nu$, as follows:
\begin{align}
& F_1(\gamma_1, \beta_1) = \sum_{\nu=1}^n
 (1-2y_\nu)
 \sum_{\mathbf{b} \in \mathbb{F}^k_2}
 \notag\\
 &
 \sum_{\mathbf{a} \in \mathbb{F}^\rho_2: \mathbf{b} = \mathbf{G}^\mathbf{b} \mathbf{a}}
 A_\nu^{\mathbf{a}, \mathbf{b}}
 (\jmath s)^\omega c^{(\rho-\omega)}
 (-\jmath s')^\varpi c'^{(d^\mathrm{c}_\nu-\varpi)}
 .
\end{align}

\subsection{Numerical Validation in Quantum Simulations}

Using the above-described method, we can systematically derive the
analytic expression of $F_1(\gamma_1, \beta_1)$ for
level-1 QAOA decoder of any arbitrary binary linear codes given its
generator matrix.  For instance, we obtain the following quantity for
(16, 5) systematic Reed--Muller codes:
\begin{align}
 &F_1(\gamma_1, \beta_1)
 = \tfrac{1}{32} \sin(4\gamma_1)  \sin(2\beta_1) \notag\\
 &
 \bigl( 4
 (\cos(4\gamma_1) + \cos(12\gamma_1) + \cos(20\gamma_1) + \cos(24\gamma_1)) 
 \sin^4(2\beta_1) \notag\\
 &+
 5
 (\cos(4\gamma_1)+\cos(12\gamma_1))
(25 + 36\cos(4\beta_1) + 3\cos(8\beta_1))
 \bigr).\notag
\end{align}

\begin{figure}[t]
 \centering
 \subfloat[(16, 5, 8) systematic Reed--Muller code: $\bar{d}^\mathrm{c}=2.50$]
 {\includegraphics[width=0.8\linewidth]{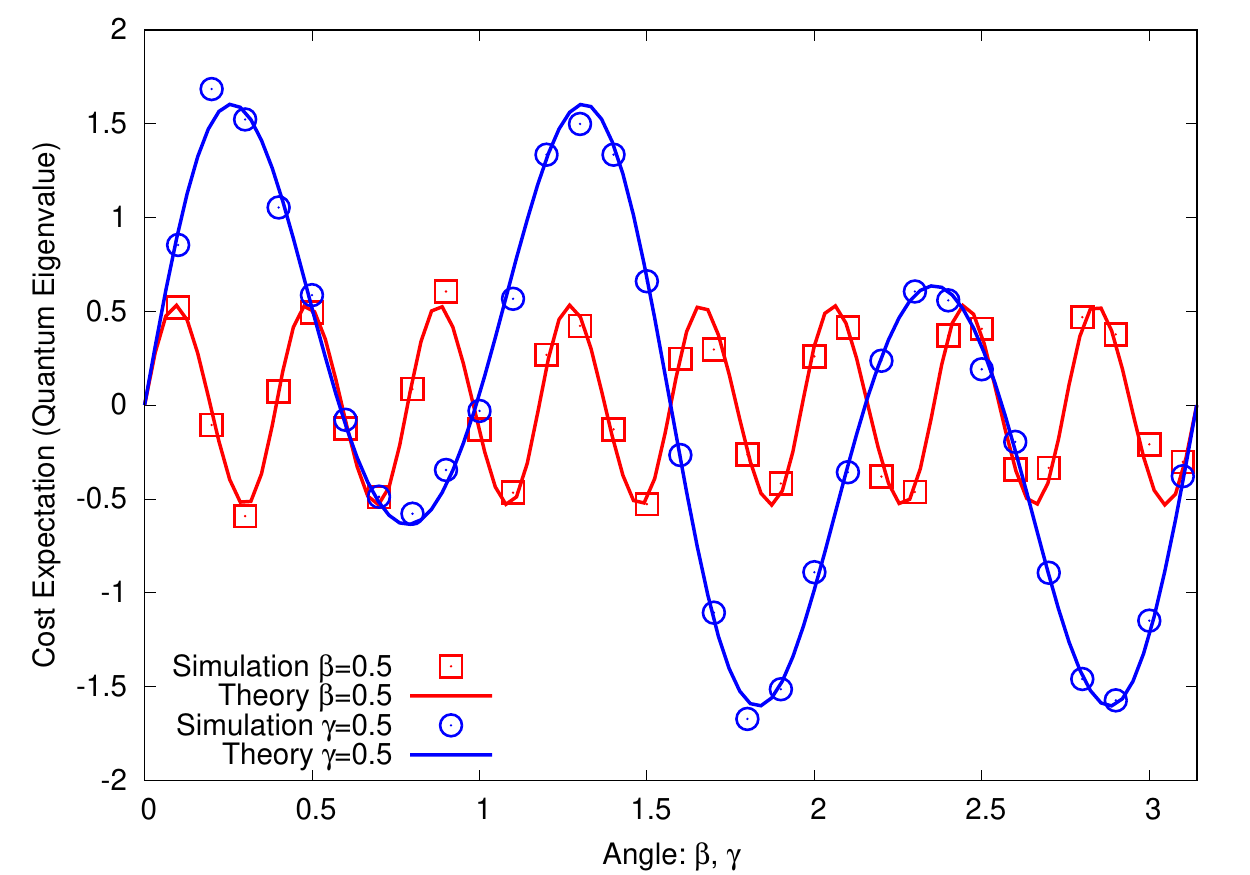}\label{fig:rm}}\\
 \subfloat[(7, 4, 3) systematic Hamming code: $\bar{d}^\mathrm{c}=1.86$]{\includegraphics[width=0.8\linewidth]{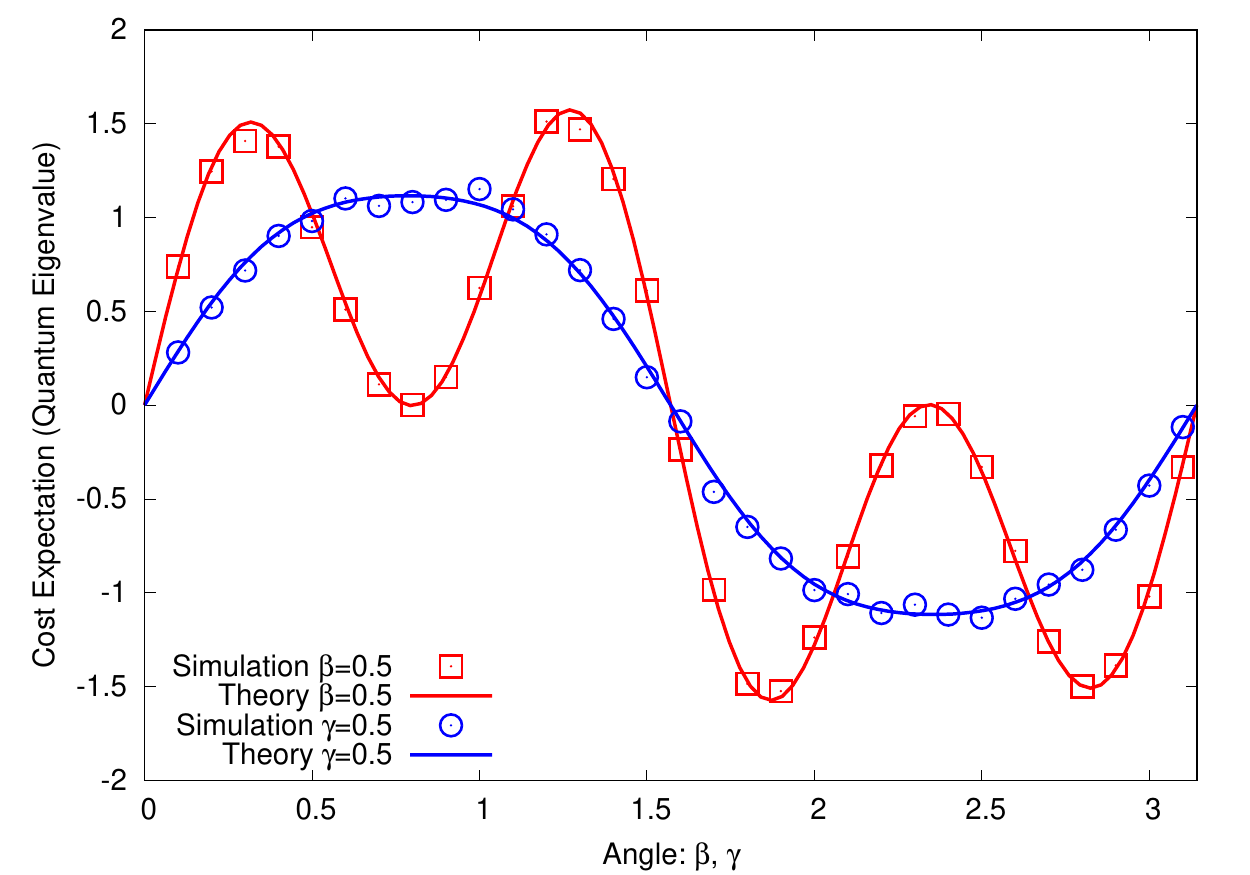} \label{fig:ham}}
 \caption{Cost expectation $F_1(\gamma_1, \beta_1)$ for level-1 QAOA decoding.}
 \label{fig:bg}
\end{figure}

Fig.~\ref{fig:bg} shows the cost
expectation $\langle C \rangle$ with sweeping angles of $\beta_1$ or
$\gamma_1$, for the (16, 5, 8) systematic Reed--Muller code. For quantum simulations,
we use qiskit to obtain the averaged cost function over $8192$-shot
measurements. It was verified that our theoretical analysis agrees well
the simulation results.

\begin{figure*}[t]
 \centering
 \subfloat[$\bar{d}^\mathrm{c}=1.71$]{\includegraphics[width=0.33\linewidth]{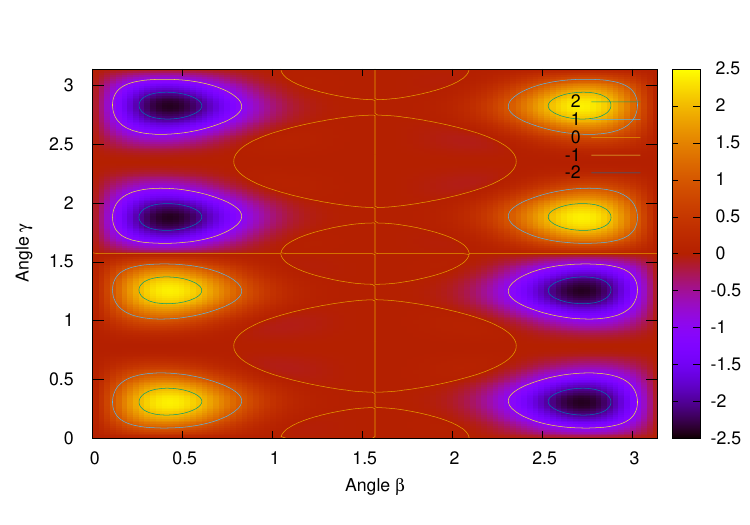}\label{fig:1.71}}
 \hfill
 \subfloat[$\bar{d}^\mathrm{c}=1.86$]{\includegraphics[width=0.33\linewidth]{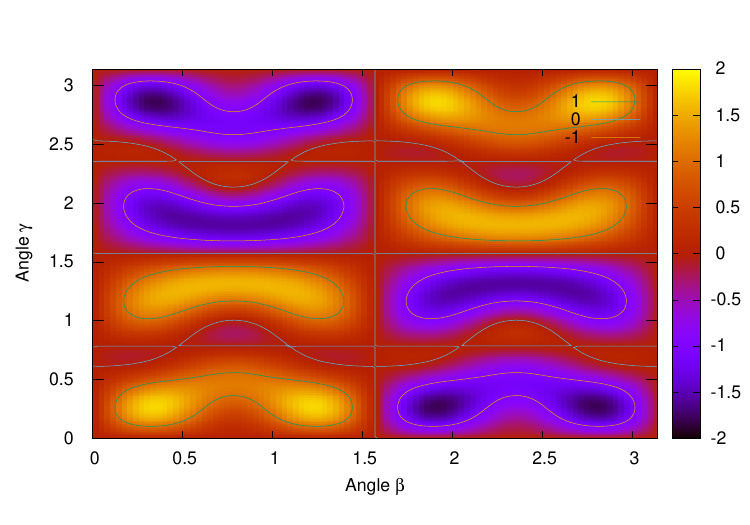}\label{fig:1.86}}
 \hfill
 \subfloat[$\bar{d}^\mathrm{c}=2.29$]{\includegraphics[width=0.33\linewidth]{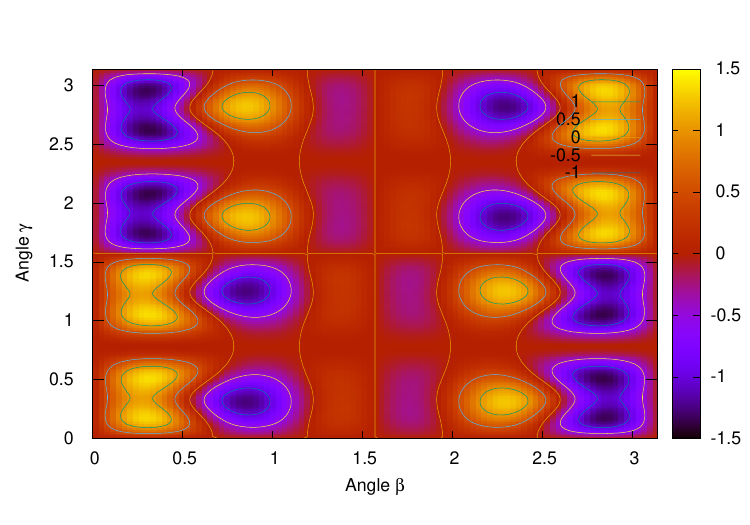}\label{fig:2.29}}
 \\
 \subfloat[$\bar{d}^\mathrm{c}=2.00$]{\includegraphics[width=0.33\linewidth]{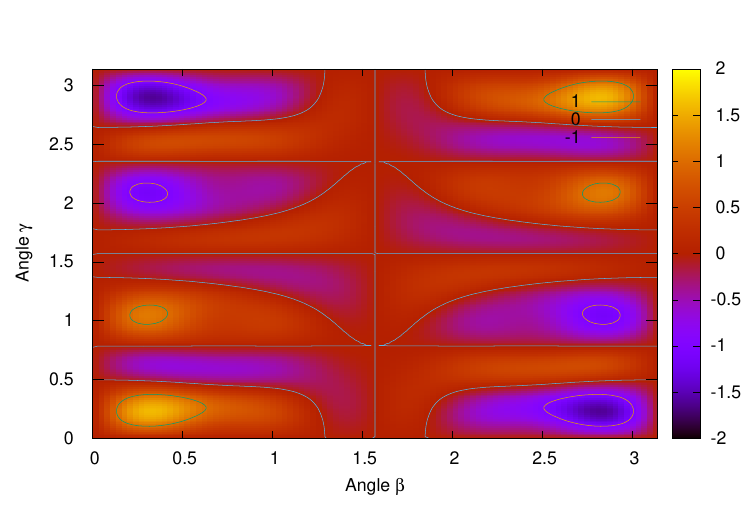}\label{fig:2.00}}
 \hfill
 \subfloat[$\bar{d}^\mathrm{c}=2.14$]{\includegraphics[width=0.33\linewidth]{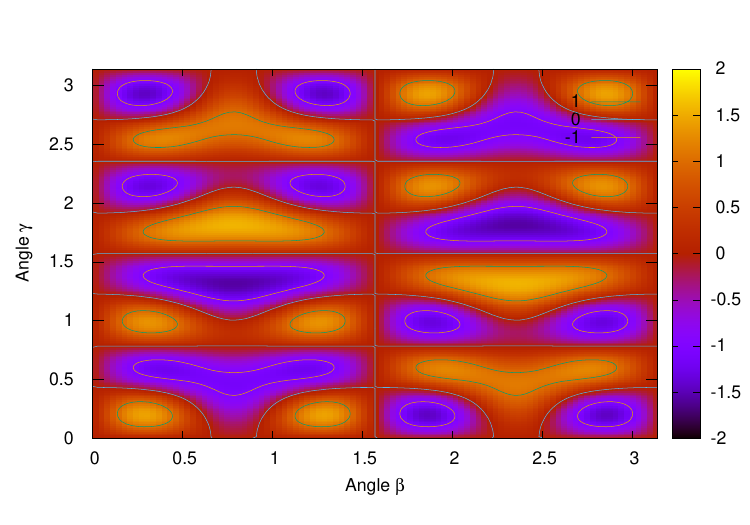}\label{fig:2.14}}
 \hfill
 \subfloat[$\bar{d}^\mathrm{c}=2.43$]{\includegraphics[width=0.33\linewidth]{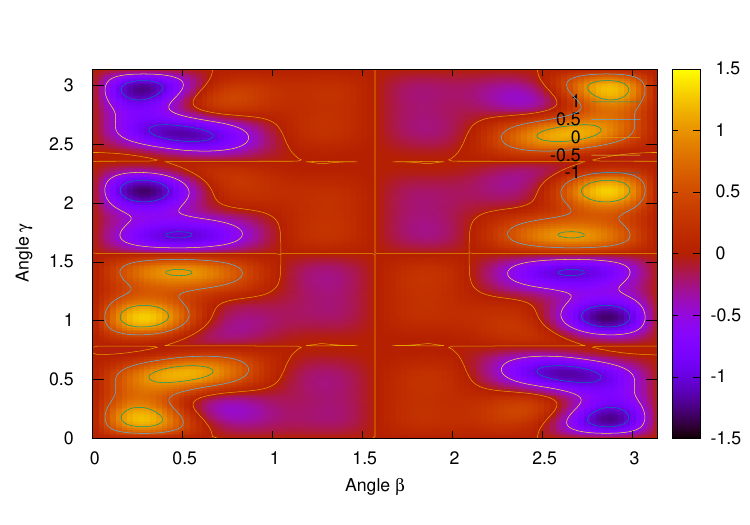}\label{fig:2.43}}
 \\
 \subfloat[$\bar{d}^\mathrm{c}=2.57$]{\includegraphics[width=0.33\linewidth]{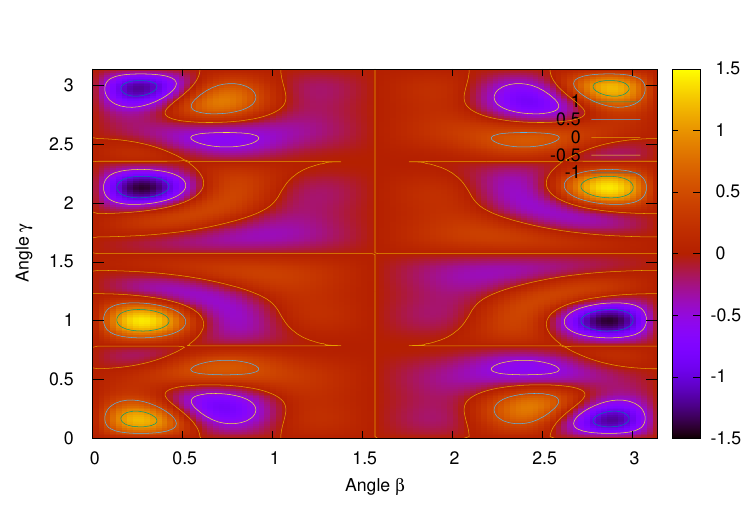}\label{fig:2.57}}
 \hfill
 \subfloat[$\bar{d}^\mathrm{c}=2.71$]{\includegraphics[width=0.33\linewidth]{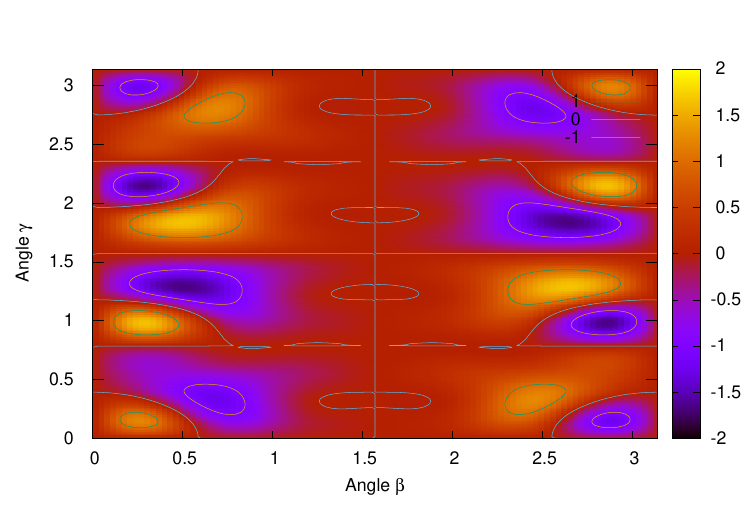}\label{fig:2.71}}
 \hfill
 \subfloat[RM code]{\includegraphics[width=0.33\linewidth]{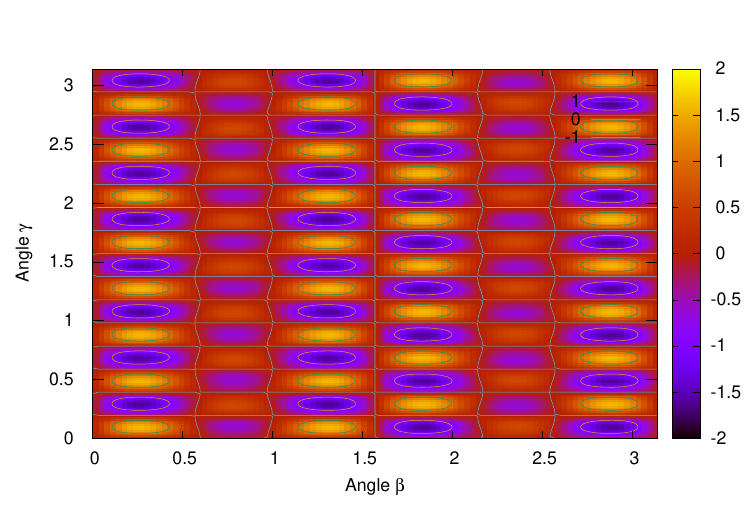}\label{fig:rm}}
 \caption{Landscape of cost expectation $F_1(\gamma_1, \beta_1)$ for level-1 QAOA decoding of (7, 4) Hamming codes.}
 \label{fig:map}
\end{figure*}

\subsection{Degree Distribution Optimization}

Figs.~\ref{fig:map}\subref{fig:1.71}--\subref{fig:2.29} show landscape
of analytic cost expectation $F_1(\gamma_1, \beta_1)$ for Hamming codes
with different generator matrix having an average degree
$\bar{d}^\mathrm{c}$ of $1.71$, $1.86$, and $2.29$, respectively. It was
shown that the quality of QAOA decoding highly depends on degree
distributions even though those basis-transformed codes have identical
Hamming weight spectrum.

Table~\ref{tab:deg} lists the theoretical cost functions derived by our
method for eight different Hamming codes with an average degree from
$1.71$ to $2.71$. From the analytical expression, we can obtain optimal
angle parameters $\gamma^\star_1$ and $\beta^\star_1$ to maximize the
cost expectation. It can be seen that the maximum cost expectation
$F_1^\star$ tends to improve as the average degree decreases. For
example, lowest-degree non-symmetric Hamming code achieves
$F_1^\star=2.409$ which is larger than na\"{i}ve random sampling, i.e.,
$\langle C \rangle=0$, whereas a smaller cost of $F_1^\star=1.790$ is
achieved by higher-degree systematic Hamming code. This trend suggests
that low-density generator-matrix (LDGM) codes can be a good candidate
for level-1 QAOA decoder. This is intuitive because there exist fewer
qubit interactions in cost Hamiltonian operator $U(C, \gamma)$.

\begin{table*}
 \centering
 \setlength{\tabcolsep}{0.8pt}
 \caption{Theoretical Expression of Level-1 QAOA Cost Expectation for Decoding (7, 4) Hamming Codes}
 \label{tab:deg}
 \begin{scriptsize}
 \begin{tabular}{|c|c|c|c|c|c|}
  \hline
 $\bar{d}^\mathrm{c}$ & $\mathbf{P}$ & $F_1(\gamma_1, \beta_1)$ & $F_1^\star$ & $\beta^\star_1$ & $\gamma^\star_1$
                      \\
  \hline
  1.71 & $\left[\begin{smallmatrix}
    1&0&0&0\\
    0&1&0&0\\
    0&0&1&0\\
    1&0&0&1\\
   \end{smallmatrix}\right]$
  & $3sc^2s'(1+c')^2-sc^2s'^3(c^2-3s^2)(c^2-s^2)$
      & 2.409 & 0.424 & 0.311 \\
  \hline
  1.86 & $\left[\begin{smallmatrix}
    1&0&0&0\\
    0&1&0&0\\
    0&0&1&0\\
    0&0&0&1\\
   \end{smallmatrix}\right]$
  & $-2sc(c^2-s^2)s'(1-3c'^2)+3sc^2s'(1+2c'^2)$
      & 1.790 & 0.345 & 0.277 \\
  \hline
  2.00 & $\left[\begin{smallmatrix}
    1&0&0&0\\
    1&1&0&0\\
    0&0&1&0\\
    0&0&0&1\\
   \end{smallmatrix}\right]$
  & $sc^2s'(1+c'+c'^2+3c'^3)+2sc(c^2-s^2)s'(1+c'^2+2c'^3)-sc^2(c^2-3s^2)(c^2-s^2)s'^3(1+c')$
      & 1.606 & 0.329 & 0.239 \\
  \hline
  2.14 & $\left[\begin{smallmatrix}
    1&1&0&0\\
    0&1&0&0\\
    0&1&1&0\\
    0&0&0&1\\
   \end{smallmatrix}\right]$
  & $3sc^2s'(c'^2-s'^2) +2sc(c^2-s^2)s'(1+5c'^2)$
      & 1.562 & 0.785 & 1.820 \\
  \hline
  2.29 & $\left[\begin{smallmatrix}
    1&0&0&0\\
    0&1&0&0\\
    0&0&1&0\\
    1&1&1&1\\
   \end{smallmatrix}\right]$
  & $-3sc^2s'(1-c'-3c'^2) +sc^2(c^2-3s^2)(c^2-s^2)s'(1+3c'+3c'^2)$
      & 1.367 & 0.310 & 0.512 \\
  \hline
  2.43 & $\left[\begin{smallmatrix}
    1&1&1&1\\
    0&1&0&0\\
    0&0&1&0\\
    0&0&0&1\\
   \end{smallmatrix}\right]$
  & $sc^2s'c'(1+2c'+3c'^2) -2sc(c^2-s^2)s'(1+c'-2c'^2-2c'^3) +sc^2(c^2-3s^2)(c^2-s^2)s'(1+3c'+2c'^2+c'^3)$
      & 1.308 & 0.283 & 1.034 \\
  \hline
  2.57 & $\left[\begin{smallmatrix}
    1&1&1&1\\
    0&1&1&1\\
    0&0&1&0\\
    0&0&0&1\\
   \end{smallmatrix}\right]$
  & $-sc^2s'(1+2c'-3c'^2-3c'^3) -2sc(c^2-s^2)s'(1-3c'^2-2c'^3) +sc^2(c^2-3s^2)(c^2-s^2)s'(1+2c'+3c'^2+c'^3)$
      & 1.420 & 0.275 & 1.005 \\
  \hline
  2.71 & $\left[\begin{smallmatrix}
    1&1&1&1\\
    0&1&1&1\\
    1&0&1&0\\
    0&0&0&1\\
   \end{smallmatrix}\right]$
  & $-3sc^2s'^3(1+c')+2sc(c^2-s^2)s'c'(1+2c')(1+c')+sc^2(c^2-3s^2)(c^2-s^2)(3+3c'+c'^2)s'c'$
      & 1.671 & 0.506 & 1.846 \\
  \hline
 \end{tabular}
 \end{scriptsize}
\end{table*}

\subsection{Higher-Level QAOA Decoder}

\begin{figure}[t]
 \centering
 {\includegraphics[width=0.8\linewidth]{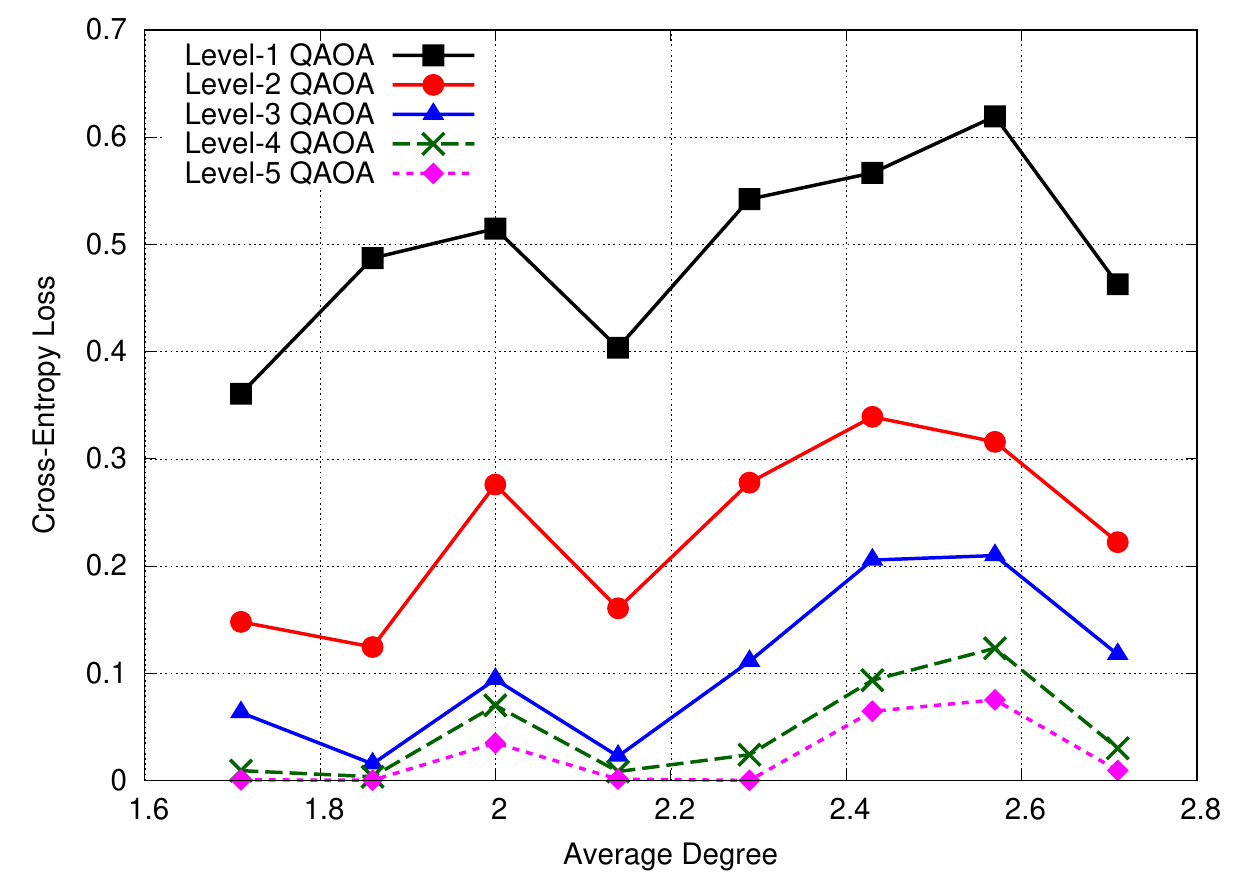}}
 \caption{Cross-entropy loss for level-$p$ QAOA decoding of Hamming codes. Angles vectors are optimized by NM method.}
 \label{fig:vqe}
\end{figure}

Fig.~\ref{fig:vqe} shows the cross-entropy loss as a function of average
degree for level-$p$ QAOA decoding of Hamming codes. The $2p$ angle
parameters were optimized by VQE employing NM method. One can see that
higher-level QAOA offers significant gain in decoding accuracy,
approaching to error-free decision, i.e., zero
cross-entropy. Interestingly, systematic Hamming code achieves the best
performance for $p\geq 2$ unlike level-1 QAOA. Systematic way to design
generator matrix for high-level QAOA is an open question to be studied
in the future.

\subsection{Success Rate of ML Decision}

\begin{figure}[t]
 \centering
 {\includegraphics[width=0.8\linewidth]{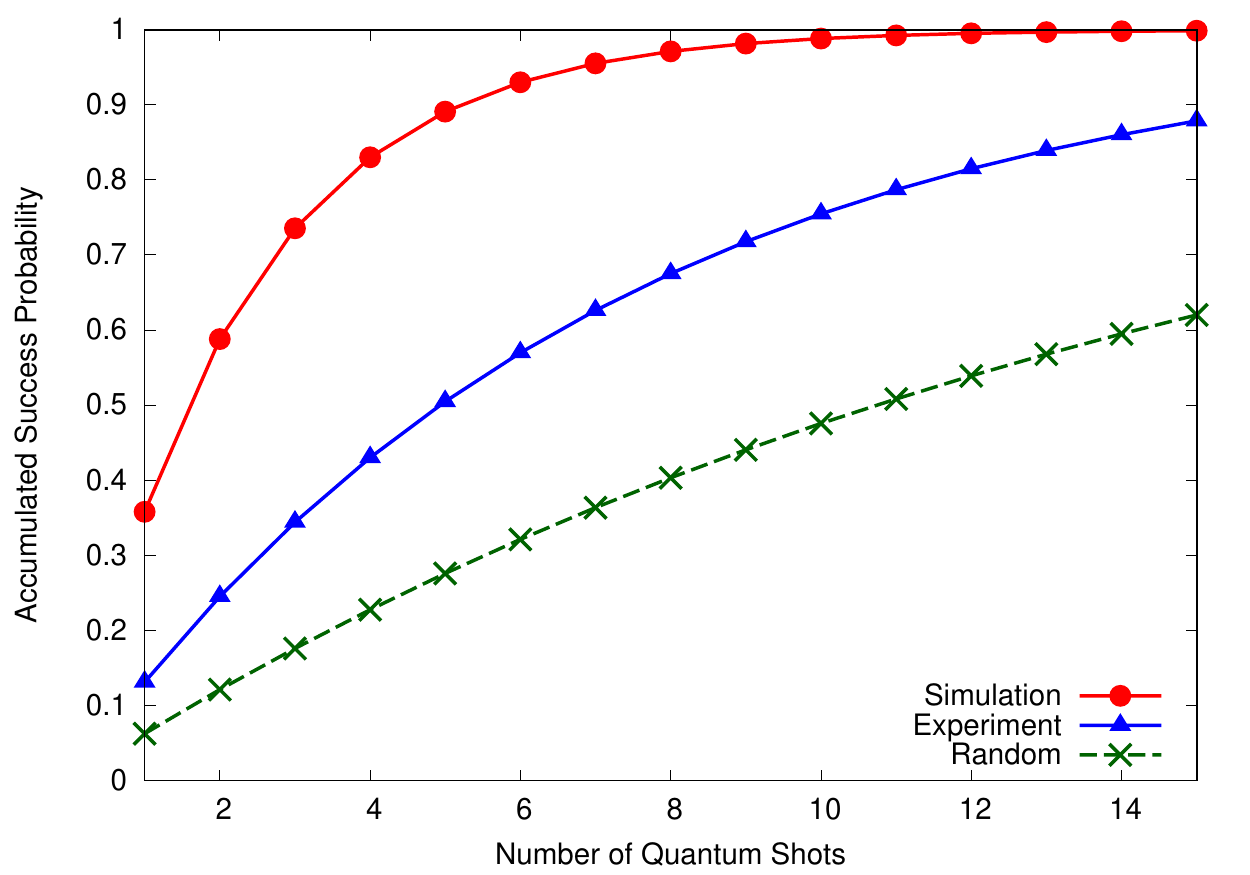}}
 \caption{ML-decision success rate of level-1 QAOA over quantum measurements for Hamming code ($\bar{d}^\mathrm{c} = 1.71$).}
 \label{fig:shot}
\end{figure}

Fig.~\ref{fig:shot} plots the accumulated success probability that QAOA
measurement gives the ML decision. For real quantum processor, we use
\textit{ibmq\_16\_melbourne}. The success rate increases with the number
of quantum shots. Although the real quantum processor has a degraded success
rate, it is still much better than na\"{i}ve random decision.

\label{sec:sim}

%



\section{Conclusion}
\label{sec:con}

We proposed to make use of QAOA algorithm for classical channel
decoding. Theoretical analysis of QAOA decoding was investigated
and insight to optimize generator matrix was provided by discussing
degree distributions of Hamming codes. How to design long codes and
dealing with soft information for high-level QAOA will be studied
in the future.


\ifCLASSOPTIONcaptionsoff
  \newpage
\fi

\bibliographystyle{IEEEtran}

\bibliography{IEEEabrv,matsumine}

\newpage

\begin{figure}[t]
 \centering
 \includegraphics[width=0.8\linewidth]{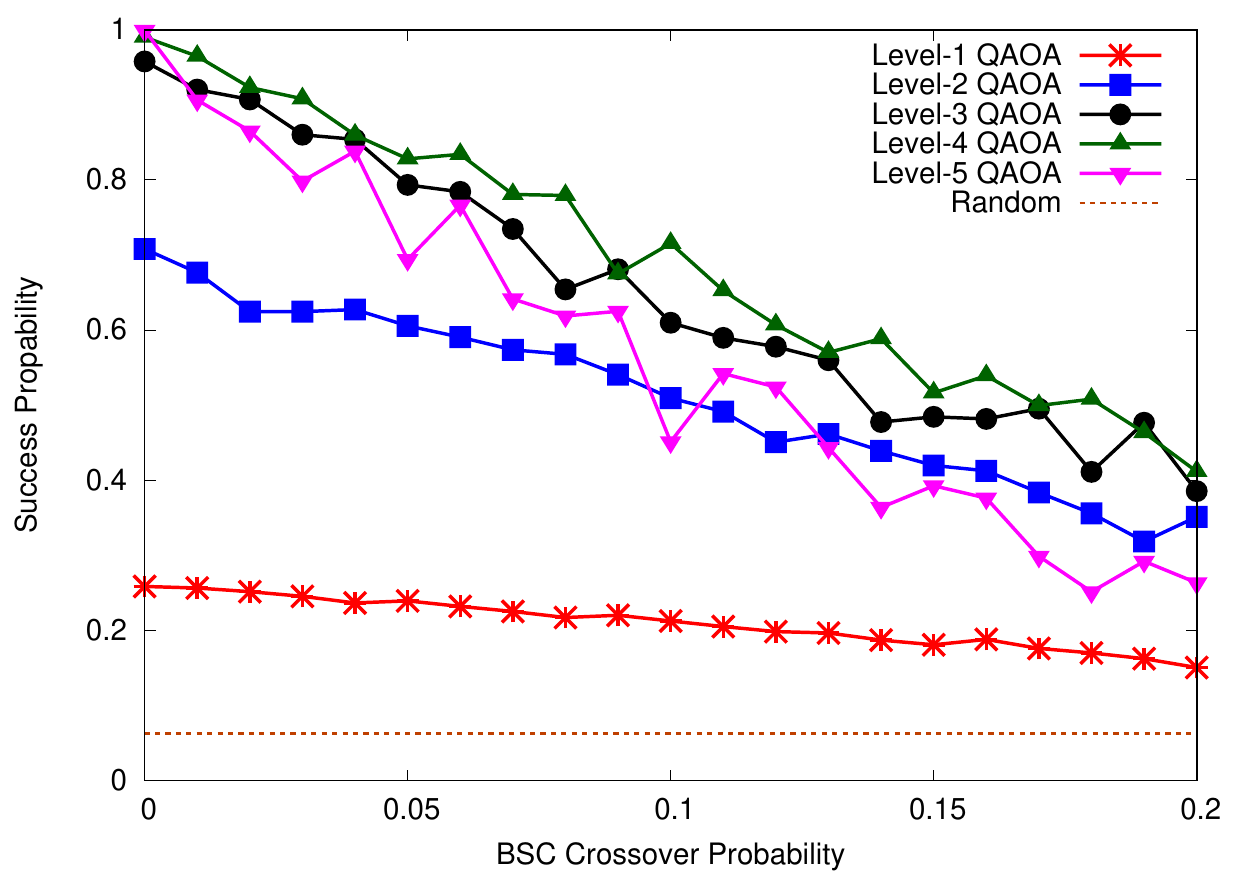}
 \caption{Single-shot QAOA decoding performance over BSC channels for Hamming codes ($\bar{d}^\mathrm{c} = 1.86$).}
\end{figure}

\end{document}